\documentclass{aa}

\usepackage{graphicx}
\usepackage{txfonts}
\usepackage{amstext}
\usepackage[colorlinks=true,allcolors=blue]{hyperref}
\usepackage{xcolor}
\hypersetup{colorlinks = true, citecolor = {blue}}

\renewcommand{\arraystretch}{1.8}

\def\gapprox{\;\rlap{\lower 3.0pt                       
        \hbox{$\sim$}}\raise 2.5pt\hbox{$>$}\;}
\def\lapprox{\;\rlap{\lower 3.1pt                       
        \hbox{$\sim$}}\raise 2.7pt\hbox{$<$}\;}

\newcommand{\be}{ \begin{equation} }
\newcommand{\ee}{\end{equation}}

\newcommand{\ben}{\begin{enumerate}}
\newcommand{\een}{\end{enumerate}}

\begin{document}
   
\title{Open Superclusters  
II. Discovery of an Inactive Supercluster Including a New  {Star} Cluster in the Solar Neighborhood }

\titlerunning{Open Superclusters II. A New Supercluster in the Solar Neighborhood}

\author{Juan~Casado
\inst{1}\fnmsep\thanks{\tt jcasadoo@hotmail.com}
\and
Yasser~Hendy
\inst{2}
\and
Nikola~Faltov{\'a}
\inst{3}
\and
Dana~A.~Kovaleva
\inst{4}
}

\institute{Facultad de Ciencias, Universitat Autònoma de Barcelona (UAB), 08193 Bellaterra, Barcelona, Spain
\and Astronomy Department, National Research Institute of Astronomy and Geophysics (NRIAG), 11421 Helwan, Cairo, Egypt
\and Department of Theoretical Physics and Astrophysics, Masaryk University, Kotlářská 2, 611 37 Brno, Czechia
\and Institute of Astronomy, Russian Academy of Sciences, 48 Pyatnitskya St, Moscow 119017, Russia}
  
\date{Received  / Accepted }

\abstract    
{The number of known star clusters, associations, and moving groups in the vicinity of the Sun has increased significantly due to the use of Gaia data.  }
{We investigated the existence and properties of open superclusters (OSC) within 500 pc of the Sun, with a particular focus on a newly identified OSC, designated HC8. }
{Using advanced Gaia-derived astrometric data and star cluster catalogs, we identify and analyze member stars of various clusters, deriving key parameters such as ages, distances, extinctions, and kinematic properties. }
{Our findings reveal that HC8, a previously unrecognized OSC, encompasses several young clusters, including the newly discovered cluster Duvia~1. We provide a detailed examination of HC8's star formation history and its spatial and kinematic characteristics.}
{ This study contributes to the growing body of evidence supporting the existence of highly populated primordial groups and enhances our understanding of the formation and evolution of star clusters in the Galaxy.}


 \maketitle

\section{Introduction}
\label{sec:intro}

It is widely accepted that most stars, especially the heavier ones, form in clusters, notably as part of embedded star clusters within the cores of giant molecular clouds (GMCs) \citep[e.g.,][]{2003ARA&A..41...57L}. These embedded clusters result from star formation processes in hierarchically structured gas clouds and often form in groups \citep[e.g.,][]{2006A&A...445..545P, 2009Ap&SS.324...83E, 2009ApJ...700..436D, 2016MNRAS.455.3126C, 2017A&A...600A.106C}. These groupings can be preserved until they emerge from their dense birth clouds, at which point they can be studied in the optical band {\citep[e.g.,][]{2021ARep...65..755C, 2023MNRAS.521.1399C, 2024A&A...687A..52C}}.
After this stage, these groupings are often referred to as "cluster complexes" or "primordial groups". We can tentatively define primordial groups as groupings of  {star clusters} that are in spatial proximity and share a common origin from the same GMC. These groups are characterized by their youth, with ages typically ranging from a few million years (Myr) to several tens of Myr, and are predominantly located in galactic spiral arms \citep[e.g.,][]{2013MNRAS.434..313G, 2016MNRAS.455.3126C}. Aside from stars in cluster cores, they usually contain large populations in cluster coronae  \citep[e.g.,][]{2003ARA&A..41...57L}. The Primordial Group hypothesis suggests that only young enough  {clusters} are associated, while older  {clusters} are isolated \citep{2022Univ....8..113C}.

 Until recently, studies of clusters groupings were limited due to insufficient accuracy and volume of astrometric data, as well as a lack of data on clusters' radial velocities (RVs). \cite{2017A&A...600A.106C} searched for groups of clusters in the Galaxy using RAVE  \citep{2006AJ....132.1645S} data combined with COCD \citep{2005A&A...438.1163K} data. They found 14 pairs, 4 groups with 3-5 members, and only one complex with 15 members.

Substantial and compelling evidence for the existence of primordial groups \citep{2019A&A...626A..17C, 2022Univ....8..113C, 2024A&A...687A..52C, 2024Natur.631...49S} has been gathered by the Gaia mission, which provides optical astrometry with unprecedented accuracy and scope \citep{2016A&A...595A...1G, 2023A&A...674A...1G, 2021A&A...649A...2L}. The number of known open clusters, associations, and moving groups in the vicinity of the Sun has significantly increased due to the use of Gaia data \citep[e.g.,][]{castrogea18, liupang19, simea19, castrogea20, 2020A&A...640A...1C, haoea22, heea22, 2023A&A...673A.114H, 2023ApJS..265...12Q, 2024A&A...686A..42H}. \cite{2024Natur.631...49S} have shown that 57\% of young clusters within 1 kpc of the Sun belong to just three cluster "families" that arise from three star-forming regions (SFRs).

In this series of articles  {we refer to clusters as groups of associated stars regardless they are gravitationally bound or unbound.} Open superclusters (OSCs) are defined as primordial groups of six or more  {clusters} \citep{2024A&A...687A..52C}. In that first article, we identified 17 OSCs in the third Galactic quadrant, encompassing at least 190 clusters. In this article, we focus on the OSCs less than 500 pc from the Sun, particularly a new OSC called HC8, which is studied in detail.

The paper is structured as follows. In Section~\ref{sec:meth}, we describe the input  {cluster} catalogs, the selection of  {cluster} member stars, and how the  {cluster} parameters, such as ages, distances, extinctions, and kinematic properties, are derived. In Section~\ref{sec:res}, we present the properties of the OSC, with particular attention given to the new  {cluster} Duvia~1 and the star formation history within the OSC. Finally, we summarize our results and provide conclusions in Section~\ref{sec:concl}.

\section{Methods }
\label{sec:meth}

\begin{figure}
\includegraphics[width=0.95\linewidth]{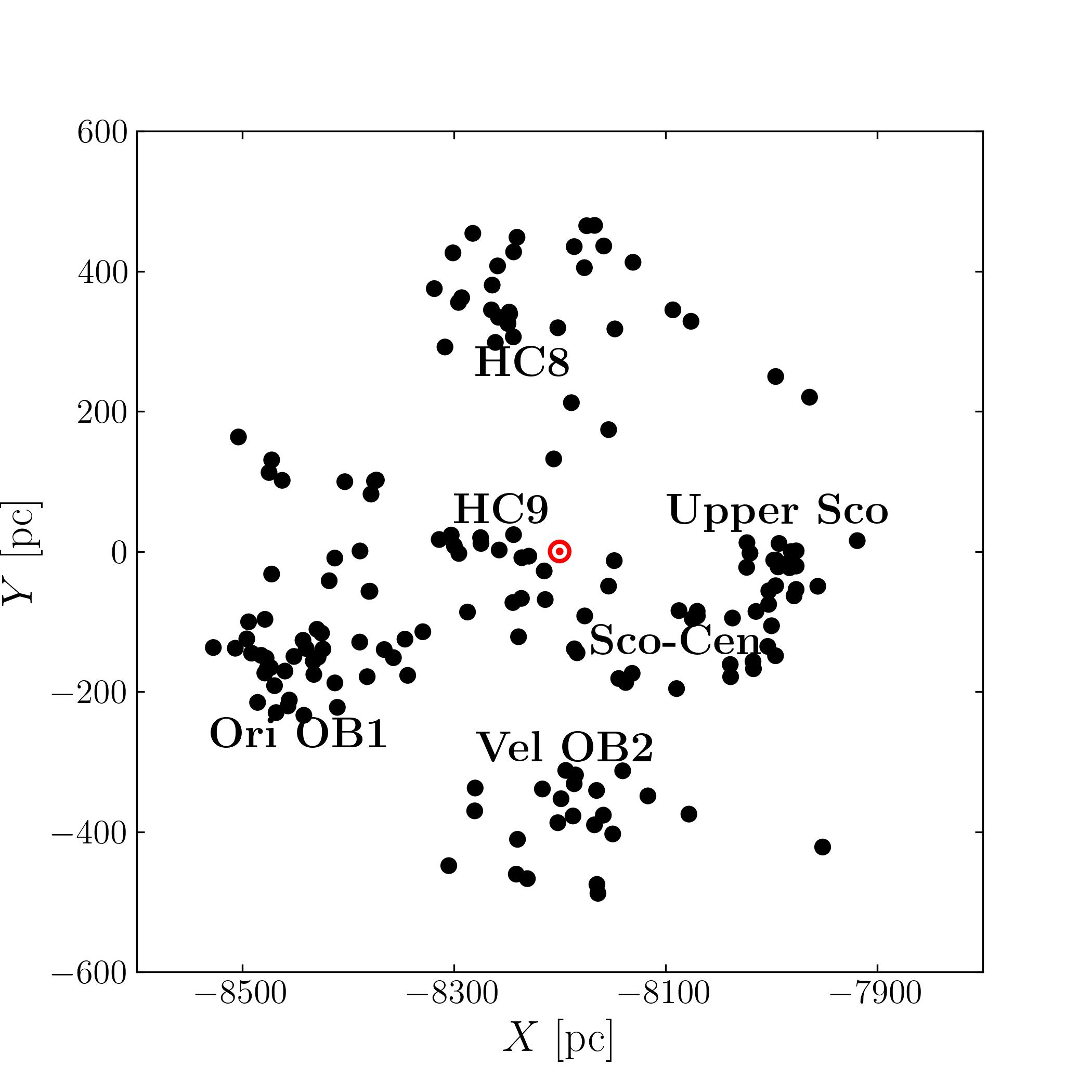}
\caption{Overdensities of open clusters $\mathrm{\le}$25 Myr old and less than 500 pc from the Sun in the HR24 catalog.  {The Sun is located at the center of the plot.}} 
\label{fig:image1}
\end{figure}

 {The identification of OSCs involves a combination of spatial, kinematic and age-related criteria. It has been performed in the Galactic coordinate frame (Galactic longitude (l), Galactic latitude (b) and distance (d)). Galactic coordinates simplify the analysis of large-scale structures in the Galactic disk. As input data,} we primarily utilized the Gaia-derived cluster catalogs of \cite{2024A&A...686A..42H}; 7167 clusters; hereafter HR24), \cite{2022ApJS..260....8H}; 541 clusters), and \cite{2020A&A...640A...1C}; 2017 clusters). These catalogs include all five astrometric Gaia measurements (position, proper motion (PM), and parallax) along with ages. To identify the member  {clusters}, we used the five average astrometric parameters of the objects in these catalogs to  {preliminary} detect local overdensities  {by eye}, as illustrated in Figure~\ref{fig:image1}.  {Then, we explored in detail the detected overdensities using a method similar to the manual search for open clusters conducted using Gaia data, as detailed in previous studies \citep{2021ARep...65..755C, 2023MNRAS.521.1399C, 2024A&A...687A..52C}.}  {This method detect groups based on spatial proximity and kinematic coherence.}
The list of candidate members was refined by discarding outliers using the following procedure: We applied criteria of a maximum radius of 150 pc and a maximum tangential velocity deviation from the mean of 10 km/s for all candidate clusters, using Eq.~\eqref{GrindEQ__1_} and Eq.~\eqref{GrindEQ__2_}, respectively.

\begin{equation} \label{GrindEQ__1_} 
\mathrm{\Delta }R=\sqrt{\left({\left(l-\overline{l}\right)}^2+{\left(b-\overline{b}\right)}^2\right)}\ \left(\frac{\pi \ 1000}{180\ \overline{plx}}\right) 
\end{equation} 

\begin{equation} \label{GrindEQ__2_} 
\mathrm{\Delta }v_{tan}=\sqrt{\left({\left(\mu -{\overline{\mu }}_{\alpha }\right)}^2+{\left(\mu -{\overline{\mu }}_{\delta }\right)}^2\right)}\ \left(\frac{\kappa}{\overline{plx}}\right) 
\end{equation}

The $\overline{l}$, $\overline{b}$, $\overline{{\mu }_{\alpha }}$, $\overline{{\mu }_{\delta }}$, $\overline{plx}$ are the means of Galactic longitude, Galactic latitude, PM in \textit{$\alpha$} and \textit{$\delta$}, and parallax, respectively. $\kappa=4.7404$ is the transformation factor from mas/yr at 1~kpc to km/s.

The likely members of HC8 have uncertainty intervals of less than 2$\sigmaup$ from the mean of the 5 dimensions. We applied enough iterations to obtain convergence with the same number of clusters from the last iteration. Next, we checked that the selected candidates have an RV spanning less than 20 km/s. The rest of the clusters in the raw list that did not pass this filtration process are also listed below as dubious candidates. It is noteworthy that we examined all OCs, both young and old, for these overdensities  {in space and PMs},  {and we consistently found only young clusters (ca. 100 Myr or less). The age consistency is considered an independent evidence for likely OSCs.} 

 {The 150 pc radius}  threshold was chosen considering the typical size of the aggregates found in the previous literature \citep{1988SvAL...14..347E, 2006A&A...445..545P, 2014ApJ...787L..15E, 2017A&A...600A.106C}. The maximum span in PMs (20 km/s) was selected because the usual difference between sibling clusters is less than 10 km/s \citep{2017A&A...600A.106C, 2021ARep...65..755C}, while the maximum RV difference between two open clusters within the same complex is approximately 20 km/s \citep{2005ApJ...630..879F}.

For RSG~7 and the new cluster Duvia~1, we selected the data from Gaia DR3 of stars with G$\mathrm{<}$18, as the \textit{plx} and PM errors increase exponentially with magnitude. We calculated age, distance, and extinction by means of the ASTECA algorithm \citep{2015A&A...576A...6P}, using the theoretical isochrones of PARSEC v1.2S \citep{2012MNRAS.427..127B} to fit the CMDs.  {The} range of ages used ($\log$ \textit{t} in years) was 6.0 to 10.15, with a step of 0.05,  {for solar metallicity Z=0.0152}. We applied the FASTMP algorithm \citep{2023MNRAS.526.4107P} to identify probable members of each cluster.

 To obtain the mass of every probable member star, we used the high degree polynomial equation between the absolute magnitudes and the masses of the main sequence stars from the isochrones of the same age and metallicity \citep{2023MNRAS.521.1399C}. The total mass is the summation of masses of each probable member star.

 We obtained the mean components of the spatial velocity of the supercluster (U${}_{0}$, V${}_{0}$, W${}_{0}$) as the mean of (U${}_{i}$, V${}_{i}$, W${}_{i}$) velocities of its component clusters. We analyzed the internal kinematics of the OSC by looking at residual spatial velocities of included clusters with respect to the mean spatial velocity of the supercluster. This procedure is realized as follows. For the clusters included in HR24, we used the coordinates, \textit{plx} and PMs of them to calculate their individual velocities (U${}_{i}$, V${}_{i}$, W${}_{i}$). Regarding RVs, we obtained median RV${}_{i}$ values based on RV of probable members from HR24 instead of using mean values provided by HR24. The reason for the preference for the median RV${}_{i}$ is to decrease the contribution of outliers.

To analyze the internal kinematics of star clusters in the supercluster, we mainly relied on the dispersion of transverse velocities, given the significant uncertainties in RVs from individual stars, as provided by Gaia \citep{2023ARep...67..938K}. We applied a modification of the convergent point method \citep{2011A&A...531A..92R, 2024A&A...686A.225I} to calculate the expected transverse velocities for their coordinates, assuming that they share the spatial velocity of a star cluster. This approach allowed us to capture the relative internal kinematics of the stars based on their residual transverse velocities.

The PM expected for the n${}^{th}$-star with coordinates (\textit{l${}^{n}$, b${}^{n}$}) if it shares spatial velocity of the i${}^{th\ }$star cluster (\textit{U${}_{i}$, V${}_{i}$, W${}_{i}$}) is determined as
\begin{equation} \label{GrindEQ__3_} 
{{\mu}}^{{\mathrm{exp\_} n,\ i\ }}_l=\frac{\left(-{\mathrm{sin} l^n\textrm{·}U_i+{\mathrm{cos} l^n\ }\ }\textrm{·}V_i\right)}{{\kappa }/{{plx}^n}}; 
\end{equation} 
\begin{equation} \label{GrindEQ__4_} 
{\textrm{µ}}^{{\mathrm{exp\_} n,\ i\ }}_b=\frac{\left(-{{\mathrm{cos} l^n\ }\mathrm{sin} b^n\textrm{·}U_i-{\mathrm{sin} l^n\ }\ }\textrm{·}{{\mathrm{sin} b^n\ }\textrm{·}V}_i+{\mathrm{cos} b^n\textrm{·}W_i\ }\right)}{{\kappa }/{{plx}^n}}, 
\end{equation} 
where 
\begin{small}
\begin{equation} \label{GrindEQ__5_} 
\left[ \begin{array}{c}
U^i \\ 
V^i \\ 
W^i \end{array}
\right]={RV}^i\textrm{·}\left[ \begin{array}{c}
{\mathrm{cos} l^i{\mathrm{cos} b^i\ }\ } \\ 
{\mathrm{sin} l^i\ }{\mathrm{cos} b^i\ } \\ 
{\mathrm{sin} b^i\ } \end{array}
\right]+\frac{\kappa {\textrm{µ}}^i_l}{{plx}^i}\textrm{·}\left[ \begin{array}{c}
-{\mathrm{sin} l^i\ } \\ 
{\mathrm{cos} l^i\ } \\ 
0 \end{array}
\right]+\frac{\kappa {\textrm{µ}}^i_b}{{plx}^i}\textrm{·}\left[ \begin{array}{c}
-{\mathrm{cos} l^i\ }{\mathrm{sin} b^i\ } \\ 
-{\mathrm{sin} l^i\ }{\mathrm{sin} b^i\ } \\ 
{\mathrm{cos} b^i\ } \end{array}
\right],                              
\end{equation} 
\end{small}
$\kappa$=4.7404, indices \textit{i }vary from cluster to cluster, and the values for the coordinates, \textit{plx}, PM and RV are as described above. 

 Hence, the residual PM components are calculated as ${\Delta}\mu{}_{l}= \mu{}_{l} -{{\mu}}^{{{exp\_} n,\ i\ }}_l$, ${\Delta}\mu{}_{b}= \mu{}_{b} - {\mu}^{{\mathrm{exp\_} n,\ i\ }}_b$. To obtain residual transverse velocity, one needs to multiply PM by ${\kappa }/{{plx}^n}$.

 {For the investigation of dynamical history of the star clusters, we employed backward orbital integration using the \textit{galpy} package \citep{2015ApJS..216...29B} in the Milky Way potential. For this test, we used the simplified model that neglects physical dimensions of the clusters and considers the backward movement of their centers based on present-day 3D coordinates (RA, Dec, \textit{plx}) and 3D velocities (PM and RV).}

\section{Results and discussion}
\label{sec:res}

\subsection{Overview of the OSCs near the sun.}
\label{sec:osc}

Identifying OSCs can be relatively straightforward by plotting young clusters from the HR24 catalog on the Galactic plane (Figure~\ref{fig:image1}). {The 25 Myr threshold was selected as a compromise to have sufficient clusters to detect the groupings while avoiding too many points that would obscure them.} The overdensities of clusters in this plot reveal both known and new OSCs. The Gould Belt is observable from Ori OB1 to Upper Sco, through Sco-Cen \citep{2024A&A...687A..52C}. All clusters in this region originate from a single SFR \citep{2024Natur.631...49S}. Recent studies have identified several groupings of clusters with common locations in 6D space near the Sun, such as the Vela-Puppis SFR, including Vel OB2 \citep{2019A&A...626A..17C, 2020MNRAS.491.2205B, 2021ApJ...923...20P}; the Orion SFR, including Ori OB1 \citep{2021ApJ...917...21S, 2023ARep...67..336V, 2024A&A...687A..52C, 2024MNRAS.534.2566S}; Upper Sco \citep{2022AJ....163...24L, 2022A&A...667A.163M, 2023MNRAS.522.1288B}, and the Perseus SFR \citep[e.g.,][]{2021MNRAS.503.3232P}.

The new OSC HC9 (Figure~\ref{fig:image1}), probably the closest of these systems to the Sun, also appears to be related to the Gould Belt. It presents an elongated structure and  {includes as candidate} members: CWNU~1129, HSC~1318, HSC~1340, HSC~1481, Theia~7, Theia~54, Theia~66, Theia~71, and Theia~93 (HR24), and OCSN~51 and OCSN~274 \citep{2023ApJS..265...12Q}. However,  {the present} study focuses on the OSC HC8.

\subsection{The new OSC HC8 }
\label{sec:HC8}

We present a newly discovered OSC within 500 pc from the Sun, named HC8 (Figure~\ref{fig:image1}) following the series of previously identified OSCs \citep{2024A&A...687A..52C}. It is located in the 2nd Galactic quadrant, in the southern hemisphere, and within the Galactic disk (at galactic latitudes from 1$\mathrm{{}^\circ}$ to -11$\mathrm{{}^\circ}$).

\begin{figure}
\begin{center}
\includegraphics[width=0.99\linewidth]{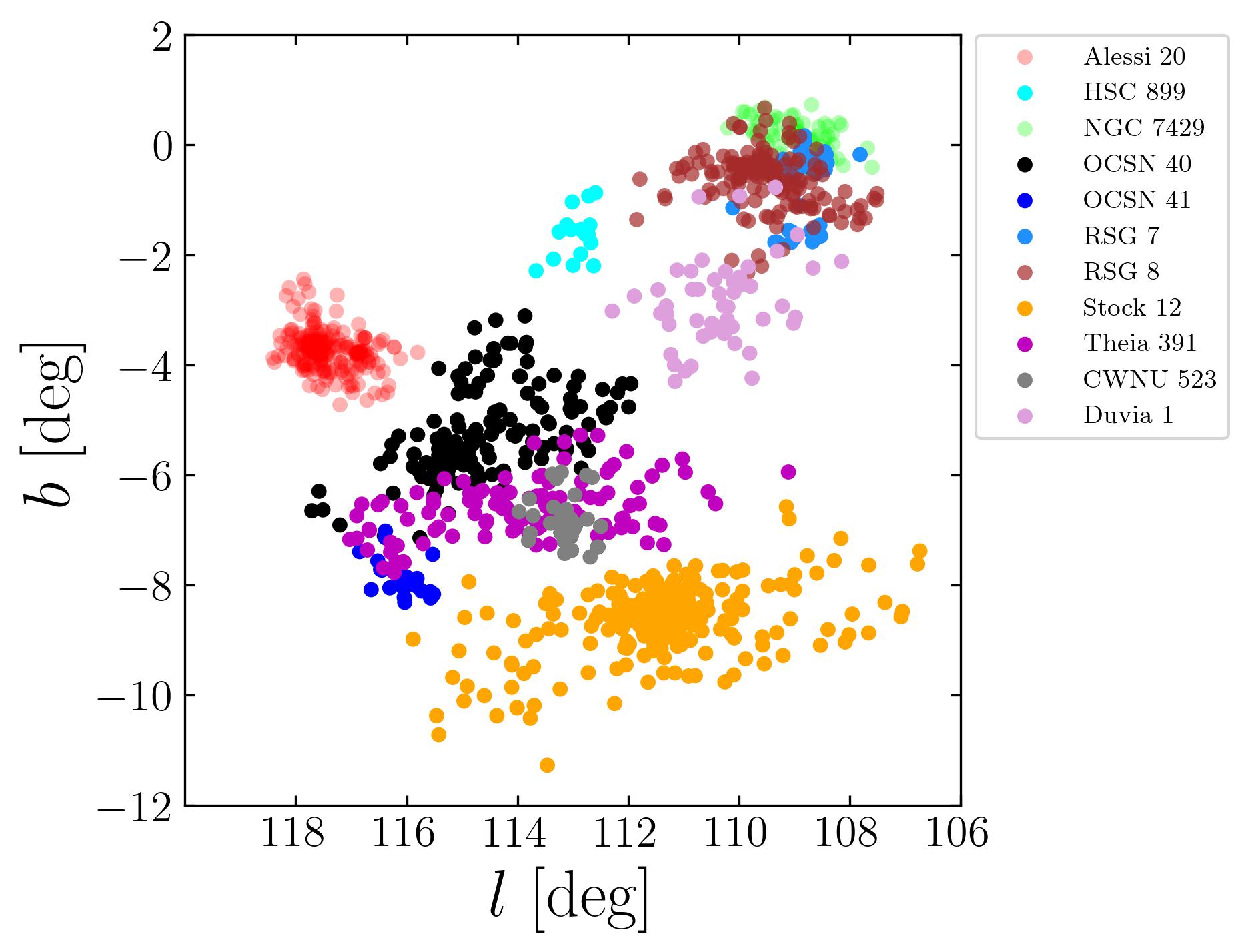}
\end{center}
\caption{Positions of the member stars of the 11 clusters identified as likely members of HC8. The members of each cluster are displayed in a different color.}
\label{fig:image2}
\end{figure}

HC8 includes ten previously known open clusters and moving groups, eight of which are listed as separate stellar aggregates in HR24, along with the newly discovered cluster Duvia 1 (see Section 3.3). Figure~\ref{fig:image2} illustrates the positions of the member stars of these 11 clusters. The characteristics of these clusters are provided in Table~\ref{tab:table1}. For RSG~7 and Duvia~1, the parameters were calculated based on lists of likely members that we compiled. For the  {rest of } clusters, parameters were obtained from  {the cited literature}.

 \cite{2021A&A...649A..54P} identified the pair RSG~7 and RSG~8 in our studied region, due to the presence of common star members in both clusters. \cite{2022ApJ...931..156P} added ASCC~127 and Stock 12 to their Alessi 20 region, attributing a filamentary structure to the system, but did not find a significant interaction between RSG~7 and RSG~8. Finally, \cite{2023ApJS..265...12Q} identified a group of three interacting clusters in the region: OCSN~40, OCSN~41, and CWNU~523.

 The list of stars for RSG~7  {in} \cite{2020A&A...640A...1C} also included stars from RSG~8 and NGC~7429. Therefore, we manually ``cleaned'' this cluster and obtained 44 likely members with G$\mathrm{<}$18 in two scattered subclusters (Figure~\ref{fig:image2}). The mean parallax is 2.31 mas, which is in good agreement with the ensemble of the Gaia literature (2.311 to 2.348 mas). The mean RV (Table~\ref{tab:table1}) encompasses the three most precise measurements (out of eight in Gaia DR3) and agrees with 4 out of 5 reported RVs in Simbad for the entire cluster.

 Of the eight clusters  of HC8 found in HR24, only three (Stock~12, Alessi~20, and RSG~8) are considered gravitationally bound open clusters. The rest are classified as mere moving groups. Three of the clusters in Figure~\ref{fig:image2} (RSG~8, Stock~12, and Theia~391) display more or less developed tidal tails stretching along the Galactic disc.

\begin{figure}
 \begin{center}
\includegraphics[width=0.99\linewidth]{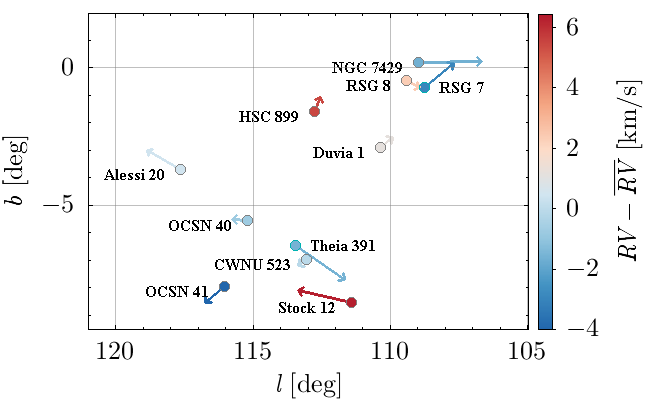}
\end{center}
\caption{Residual velocities of clusters in HC8 relative to its mean velocity.  Arrows represent direction and magnitude of the residual tangential velocities; color represents direction and respective magnitude of the residual RVs. The scale for tangential velocities is 2~km/s per arrow of 1~deg.}
\label{fig:image3}
\end{figure}

 Figure~\ref{fig:image3} provides a 3D representation of how the supercluster's members are generally moving away from each other. The new cluster Duvia~1 (Section 3.3) shares this trend.

\begin{figure*}
\begin{center}
\includegraphics[width=1\linewidth]{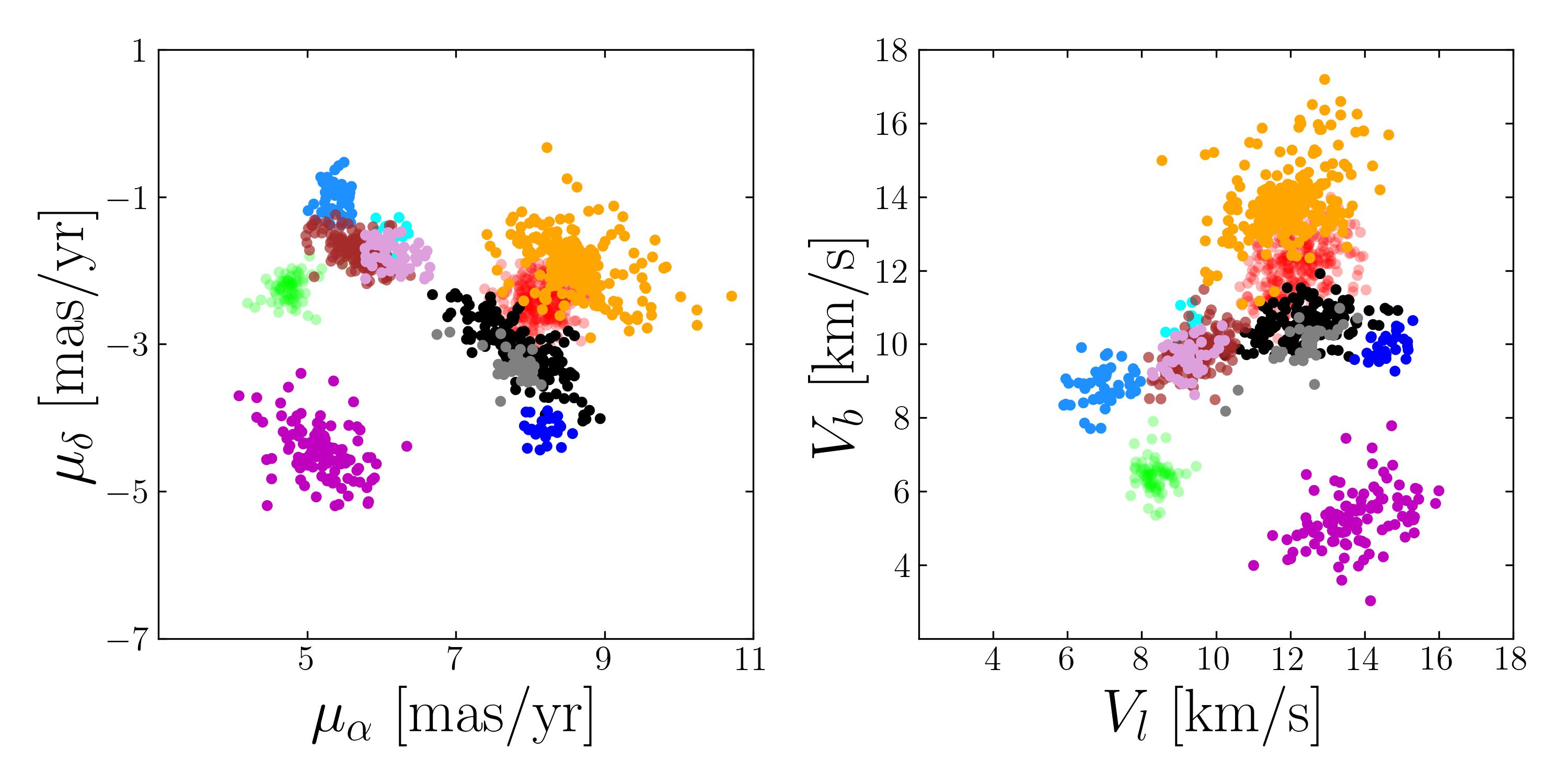}
\end{center}
\caption{ {PMs (left panel) and tangential velocities (right panel, Galactic coordinates) of the star members in the supercluster HC8. The color for each cluster is the same as in Figure~\ref{fig:image2}.}}
\label{fig:image4}
\end{figure*}

 From the range of parallax values obtained for HC8 (Table~\ref{tab:table1}), we estimate its radial size to be approximately 0.11 kpc. The radial size derived from photometric distances (d${}_{phot}$) would be 0.15 kpc. The radial size of 0.11 kpc is very similar to our estimations of transversal size (0.10 kpc). Accordingly, neither filamentary nor other obvious spatial structures are observed in HC8 \citep[Figure~\ref{fig:image2}; see, however,][]{2022ApJ...931..156P}. The span in RV of 14 km/s also matches our estimated span in tangential velocity (16 km/s) obtained from the PMs of the clusters' stars (Figure~\ref{fig:image4}). All these values are typical of OSCs \citep{2024A&A...687A..52C} and fit comfortably within the constraints detailed in Section~\ref{sec:meth}.

\begin{figure}
 \begin{center}
\includegraphics[width=0.92\linewidth]{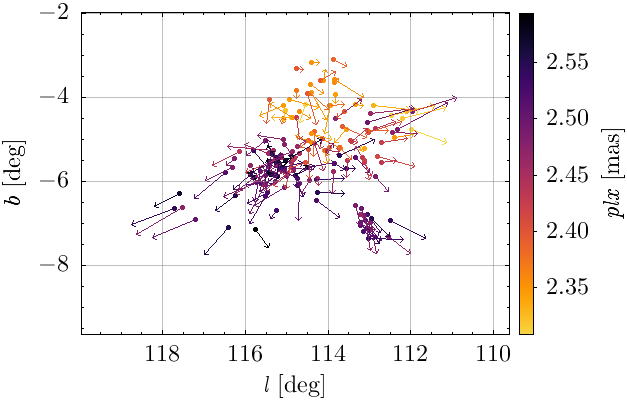}
\end{center}
\caption{Residual tangential velocities relative to the tangential velocity of OCSN~40 according to HR24. Arrows mark direction of movement and magnitude of velocity, color marks \textit{plx} according to the auxiliary axis to the right. The scale for velocities is 1 km/s per 1$\mathrm{{}^\circ}$.}
\label{fig:image5}
\end{figure}

 {Let's consider now the case of  OCSN 40 and CWNU 253. }It is noteworthy that members of OCSN~40 listed in HR24 include many sources of CWNU~523 \citep{2022ApJS..260....8H}. Figure~\ref{fig:image5} illustrates the dispersion of individual stars in OCSN~40 and CWNU~523 (clump below). The mean residual motion of CWNU~523 differs from that of the stars in OCSN~40. However, it aligns with the general trend of cluster members drifting apart from the center. This trend is not unique to this case but it is also observed in Alessi~20 and OCSN~41 (not shown). The varied \textit{plx} and the eccentric core of OCSN~40 suggest the presence of diverse subclusters, a pattern also noted in RSG~7 and Alessi~20.

 { To ascertain the possible binarity of  OCSN 40 and CWNU 253,} the estimate of the specific energy of the pair,  {using the simplified model of star clusters as attracting point masses without accounting for the external galactic field},  {is calculated from the equation} \citep[see, e.g.,][]{2007faa..book.....V}:
\begin{equation} \label{GrindEQ__6_} 
E=\frac{\Delta v^2}{2}-\frac{GM}{s}, 
\end{equation} 
where $\mathrm{\Delta}\textit{v}\mathrm{\approx}$~0.7~km/s  is our estimate for the difference in velocities of the clusters, \textit{M}$\mathrm{\approx}$100 M${_\odot}$ is our estimate for their total mass (matching the mass of OCSN~40 in HR24, 98 $\mathrm{\pm}$ 20 M${_\odot}$), and \textit{s}$\mathrm{\approx}$17.5 $\mathrm{\pm}$~0.8~pc is the distance between their centers. \textit{E} is definitely positive  {since its }negative component is smaller than the positive one by up to an order of magnitude. Although  {simplifying each cluster as a material dot at its centre of gravity} is rough taking into account that the linear scale of OCSN~40 is comparable to the distance to CWNU~523, the conclusion that CWNU~523 is not gravitationally bound to OCSN~40 aligns with the kinematics of its star members (Figure~\ref{fig:image5}).

 If the system is indeed unbound, this would endorse the notion that OCSN~40 and CWNU~523 are distinct clusters. Moreover, HR24 classifies OCSN~40 as a moving group, which  {supports} the hypothesis of distinct subclusters within OCSN~40, since unbound clusters are prone to disintegrate forming diverse subclusters  \citep[e.g. Fig~6 in][]{2024A&A...687A..52C}.

 Less likely members could include ASCC~123, ASCC~127,  {CWNU 1055,} LISC 3534, Theia~322, Theia~446, Theia~1232, OCSN~28, OCSN~36, HSC~873, HSC~895, and HSC~931 (HR24). Consequently, the total number of siblings in HC8 could increase up to  {24}.
 
\subsection{A new cluster in HC8: Duvia~1 }
\label{sec:duvia1}
 
\begin{figure*}
\begin{center}
\includegraphics[width=0.99\linewidth]{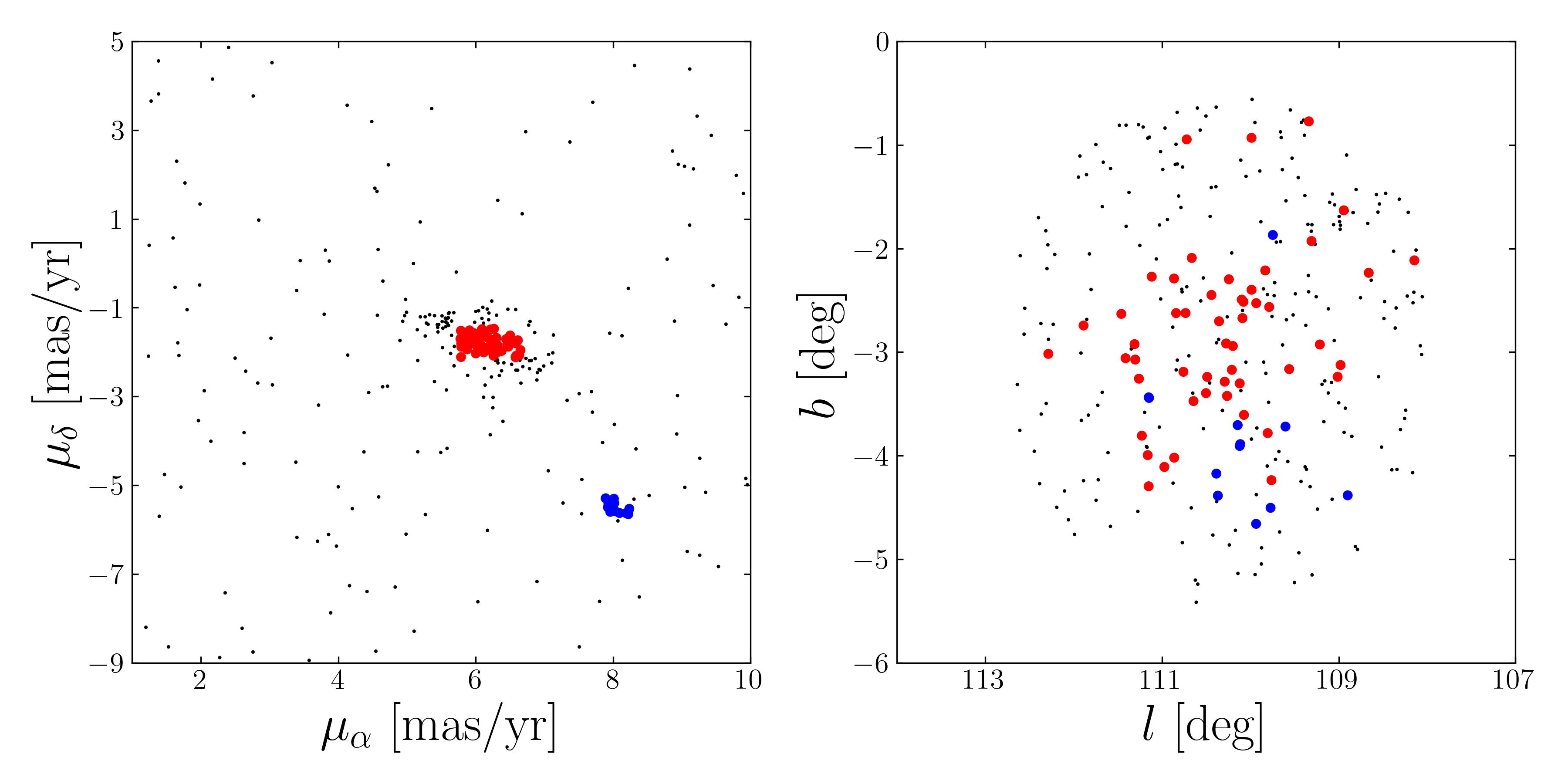}
\end{center}
\caption{ {Left panel: Vector point diagram of the likely members of Duvia~1 (red color). Another moving group lies on the below right part of the graph (blue color). Right panel: Positions of the likely members of Duvia~1  and the other moving group. The \textit{plx} cut is 2.23-2.35 mas.}}
\label{fig:image6}
\end{figure*}

 We have discovered the cluster Duvia~1 in the region of the OSC HC8 { as a striking overdensity in the vector point diagram (Figure~\ref{fig:image6}).} The mean coordinates (J2016.0) are RA=$348.81^{\circ}$ and Dec=57.64$^\circ$. 

There is at least one additional moving group (located below right in Figure~\ref{fig:image6}), consisting of a dispersed clump of only 12 stars (G$\mathrm{<}$18), but with a coherent CMD (not shown). The constraints that define the likely members of this subgroup are summarized in Table~\ref{tab:table1} with uncertainties encompassing the 12 members. The \textit{plx} distance is practically the same as that of Duvia~1 (437 $\mathrm{\pm}$ 11 pc, see below).

\begin{figure}
 \begin{center}
\includegraphics[width=0.8\linewidth]{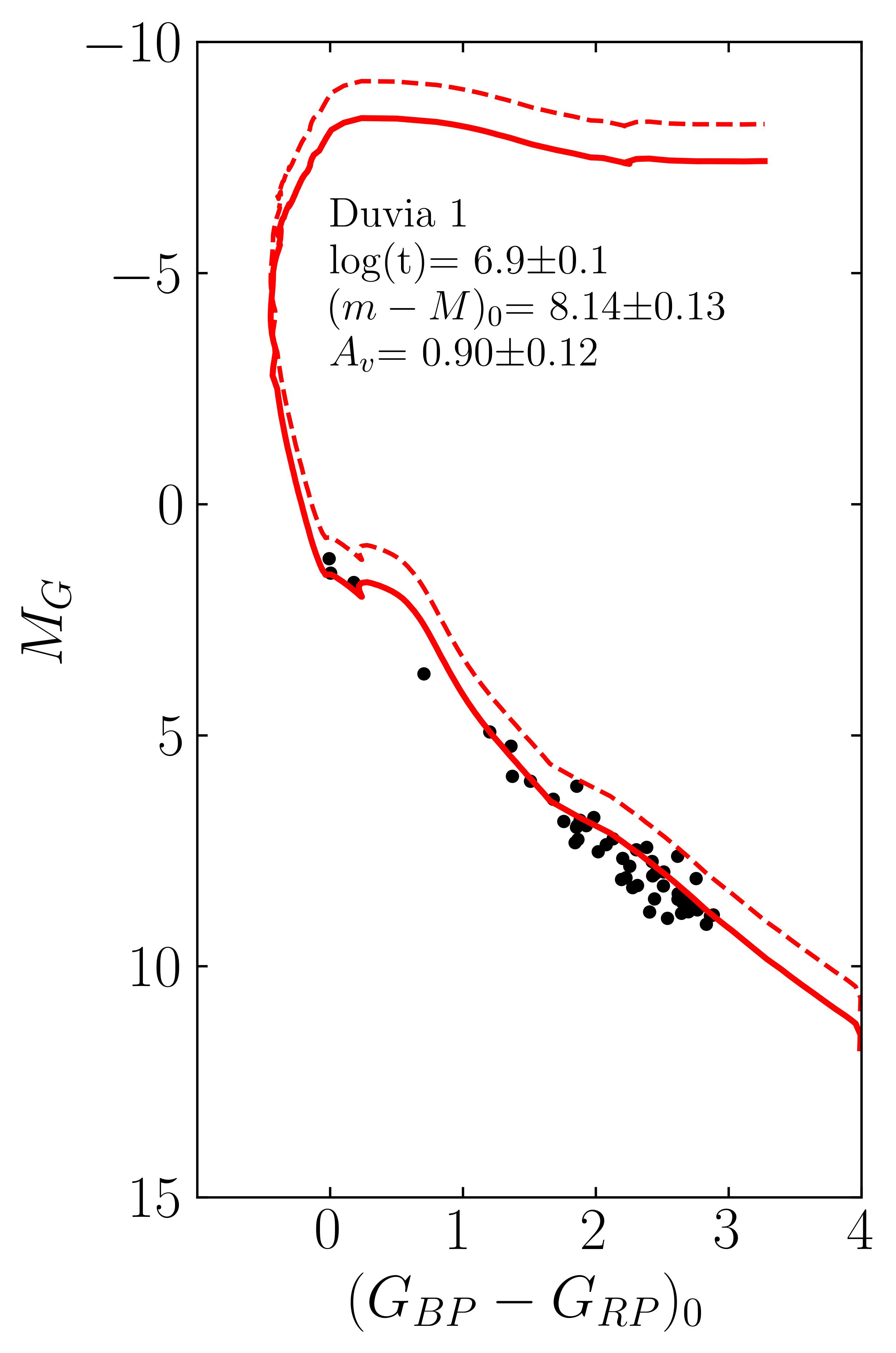}
\end{center}
\caption{ {The dereddened} CMD of the new cluster Duvia~1. The continuous line is the PARSEC isochrone fitted to our data, while dashed line represents the isochrone vertically shifted by $\mathrm{-}$0.75 mag (the locus of unresolved binaries of equal-mass components). }
\label{fig:image7}
\end{figure}

 The isochrone fitting of Duvia~1, which includes  {53} likely members, is presented in Figure~\ref{fig:image7}. From this fitting, an age of  {8$\mathrm{\pm}$2} Myr is estimated, aligning with some of the HC8 members. From the 27 likely stars of Duvia~1 within the StarHorse catalog \citep{2022A&A...658A..91A}, we found a  {median} metallicity  \textit{Z} = 0.012 $\mathrm{\pm}$ 0.003. Thus, the metallicity from the StarHorse catalog is very close to the solar metallicity.

 The distance modulus \textit{(m$-$M)${}_{0}$} is 8.14 mag, corresponding to a photometric distance \textit{(d${}_{phot}$)} of 425 pc. Fifty likely member stars in this cluster are recorded in the catalog by \cite{2021AJ....161..147B}. The median photogeometric Bayesian distance to these stars is 429 ${\pm}$ 9 pc. The mean distance of the 27 likely members cross-matched with the StarHorse catalog is 430 $\mathrm{\pm}$ 10 pc. From the median  {parallax} of 45 likely members with  relative uncertainty \textit{plx${}_{err}$/plx} $\mathrm{<}$10\% and RUWE $\mathrm{<}$1.4, a distance of 434 $\mathrm{\pm}$ 8 pc is derived. Thus, the four distances are consistent, and it is safe to state that Duvia~1 is at a heliocentric distance of 0.43 kpc, within the range of its siblings in HC8. The maximum cluster member's distance to the mean position \textit{(r${}_{max}$) }of Duvia~1 is 18 pc. Regarding the RV, there are seven likely members with reliable measurements in Gaia DR3. The median value of these measurements is --9.4~$\mathrm{\pm}$~9~km/s, which is in good agreement with the rest of the HC8 members. The main fundamental parameters are summarized in Table~\ref{tab:table1}.  {The list of probable members of Duvia~1 is provided in Table~\ref{tab:A} in the Appendix.}
 
 The extinction for Duvia~1 is  {0.90 $\mathrm{\pm}$ 0.12} mag. We also obtained values from the 3D dust sky maps: SKYMAPS \citep{2019ApJ...887...93G}, GALExtin \citep{2021MNRAS.508.1788A}, and STILISM \citep{2019A&A...625A.135L} for the new cluster. These values are 1.25 $\mathrm{\pm}$ 0.1 mag, 1.29 $\mathrm{\pm}$ 0.1 mag, and 0.96 $\mathrm{\pm}$ 0.2 mag, respectively. The 27 cross-matched likely members from the StarHorse catalog have a median of 0.91 $\mathrm{\pm}$ 0.3 mag. Thus, our estimated value is \textit{compatible with those from SKYMAPS and GALExtin, and close to those of  STILISM and the StarHorse catalog.}

\begin{table*}[tbp]
\caption{Mean parameters of the likely members of the OSC HC8, including the new cluster Duvia~1, and the overall ranges of them for the whole supercluster (last row).} 
\label{tab:table1}
\small
\setlength{\tabcolsep}{2pt}
\renewcommand{\arraystretch}{1.2}
\begin{center}
\begin{tabular}{lcccccccccc}
\hline 
{Cluster/OSC} & {\textit{l}} & {\textit{b}} & {plx} & $\mu^*_\alpha$ & $\mu_\delta$ &  d${}_{phot}$ & r${}_{max}$ & log t& RV & Ref. \\ 
(Number of stars) &{deg} &{deg} & {mas} & {mas/yr} & mas/yr & kpc & deg &  & km/s &  \\ \hline 
{OCSN~41}(27) & {116.02} & {-7.95} & {2.47 } & {8.21 } & -4.16  & 0.40 & 1.00 & 6.84 & -17 & HR24 \\ 
OCSN~40 (169) & {115.22} & {-5.55} & {2.46 } &{7.92 } & -3.13  & 0.40 & 3.47 & 6.97  & -13 & HR24 \\  
{CWNU~523}(36) & {113.13} & {-6.88} & {2.51 } & {7.89 } & -3.31  & 0.33 & - & 7.2 & -11${}^{a}$ & \cite{2022ApJS..260....8H} \\ 
{Alessi  20 }(252) & {117.64} & {-3.70} & {2.34 } & {8.12 } & -2.43  & 0.42 & 1.84 & 6.83 & -9.3 & HR24 \\  
{NGC~7429}(67) & {108.97} & {0.20} & {2.37 } &{4.77} & -2.24  & 0.42 & 1.50 & 7.89 & -12 & HR24 \\ 
{Stock 12}(273) & {111.41} &{-8.54} & {2.29 } & {8.57 } & -1.92  & 0.43 & 4.78 & 8.09 & -3.2 & HR24 \\ 
{Theia~391}(112) & {113.46} & {-6.47} & {2.21 } & {5.20 } & -4.46  & 0.45 & 4.36 & 8.16 & -13 & HR24 \\
{HSC~899}(16) & {112.77} & {-1.59} & {2.15 } & {6.03 } & -1.56  & 0.46 & 1.13 & 6.84 & -8.3 & HR24 \\ 
{RSG~8}(151) &{109.42} & {-0.46} & {2.06 } & {5.66 } & -1.72  & 0.48 & 2.58 & 7.13 & -9.7 & HR24 \\ 
{RSG~7}(44) & {108.80$\mathrm{\pm}$0.33} & {-0.73  $\mathrm{\pm}$0.61} & {2.31  $\mathrm{\pm}$0.08} & {5.37 $\mathrm{\pm}$0.15} & -1.01 $\mathrm{\pm}$0.22 & 0.41  $\mathrm{\pm}$0.03 & 1.55  & 7.2 $\mathrm{\pm}$0.1 & -14 $\mathrm{\pm}$4 & This work \\ 
Duvia~1 (53) & 110.32 $\mathrm{\pm}$0.82 & -2.85 $\mathrm{\pm}$0.77 & 2.28\newline $\mathrm{\pm}$0.04 & 6.18\newline $\mathrm{\pm}$0.26 & -1.80 $\mathrm{\pm}$0.18 & 0.43 $\mathrm{\pm}$0.04 & 2.43  & 6.9 $\mathrm{\pm}$0.1 & -9.4 $\mathrm{\pm}$9 & This  work \\ 
Clump (12)& 110.4 & -4.2 & 2.29 $\mathrm{\pm}$0.06 & 8.21 $\mathrm{\pm}$0.33 & -5.52 $\mathrm{\pm}$0.30 & ${}^b$ & 1.33 & ${}^b$ & ${}^b$ & This work \\ 
OSC HC8 (1212)  & 106.7...118.4 & -11.3...0.7 & 2.0...2.6 & 4.1...10.7 & -5.9...-0.3 & 0.33...0.48 & 1.00...4.78 & 6.83...8.16 & -3.2...-17 & This work \\ \hline
\end{tabular}
\end{center}
${}^{a}$UCC catalog (https://ucc.ar/).

${}^{b}$too few members.
\end{table*}

\begin{figure}
 \begin{center}
\includegraphics[width=0.88\linewidth]{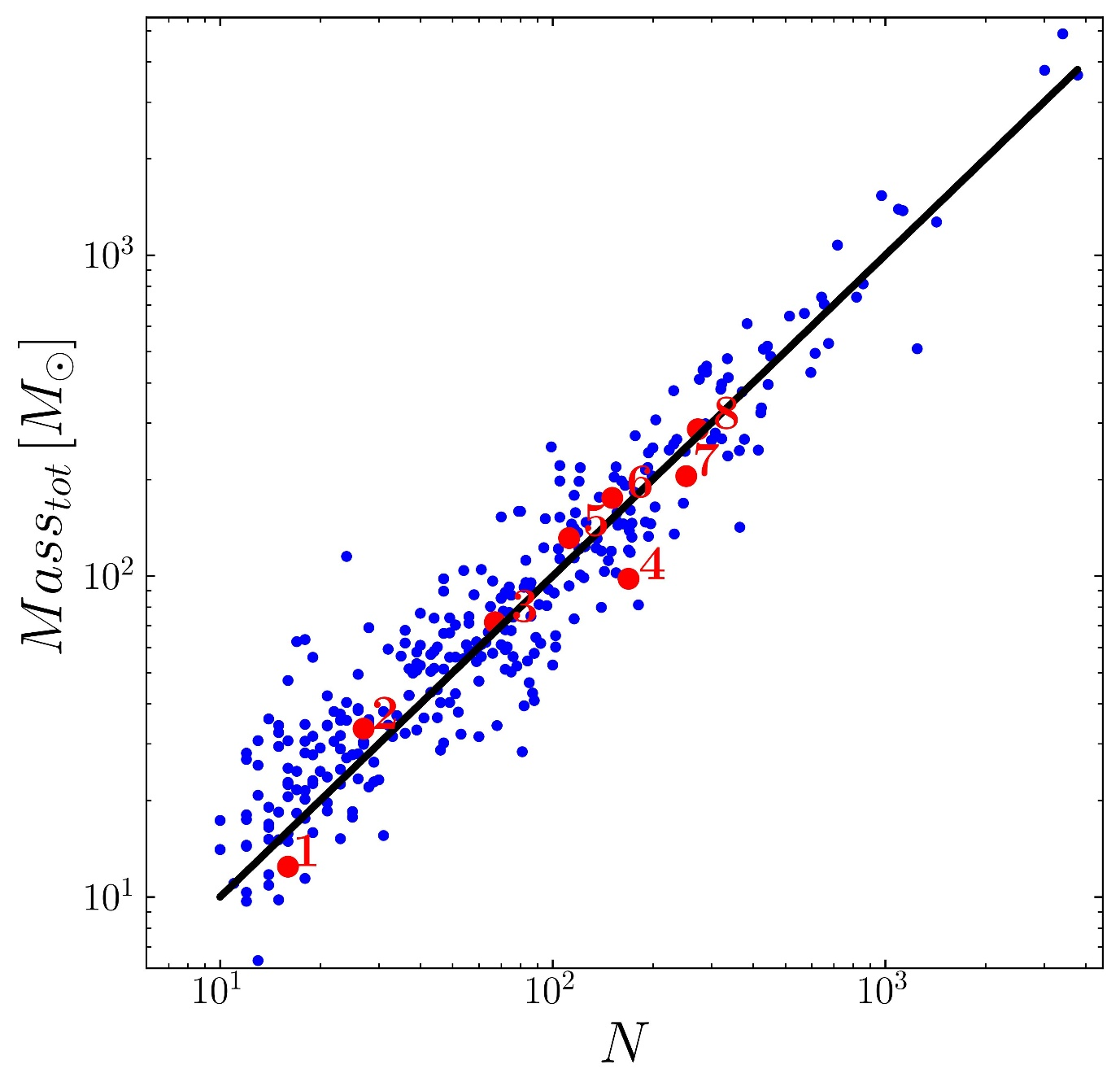}
\end{center}
\caption{The relation between the number of member stars and the total mass of the clusters in HR24 having \textit{plx} 2 to 3 mas. The red circles indicate eight probable members of HC8 (1: HSC~899, 2: OCSN~41, 3: NGC~7429, 4: OCSN~40, 5: Theia~391, 6: RSG~8, 7: Alessi 20, and 8: Stock 12).}
\label{fig:image8}
\end{figure}

 Disregarding the member stars with G$\mathrm{>}$18 and the unresolved binaries, the estimated mass of Duvia~1 is 30 M${_\odot}$.  When plotting the total mass of clusters in HR24 against the number of members (N), we observe the expected correlation: the mass is roughly proportional to N (Figure~\ref{fig:image8}). It is crucial to select the relevant \textit{plx} range because,  {perhaps} due to a systematic error, the masses of clusters in HR24 increase significantly with their distance, which is unphysical. In our case, the average mass per member is close to 1 M$_\odot$. The eight clusters of HC8 are highlighted in red. If Duvia~1 follows the same trend, its total mass  {from Figure 8} would be approximately 50 M${_\odot}$.

 These estimates allow us to address the question of the dynamical state of Duvia~1. For a stellar system in equilibrium, the virial theorem provides an approximate mass required for the system to be bound   { \citep[see, e.g., discussion in][]{2010ARA&A..48..431P}} :

\begin{equation} \label{GrindEQ__7_} 
M\cong \frac{v^2\times R}{G}\ ,             
\end{equation} 

 \noindent where \textit{v} is the dispersion of velocities in the system, \textit{R} is a characteristic radius, and \textit{G} is the gravitational constant.
 
 For Duvia~1, we estimate the  {3D} dispersion of velocities to be 0.58 km/s, and the characteristic radius (based on the half-distance between  {first quartile (Q1)} and  {third quartile (Q3)} of the distribution of stars in space) to be 9.2~pc. This leads to  {roughly} estimated bound mass M$\mathrm{\approx}750 M{}_{\mathrm{\odot }}$, which exceeds our estimate for its actual mass (50 M${}_{\mathrm{\odot }}$)
by  {at least }an order of magnitude. Thus, Duvia~1 is clearly an unbound cluster, i.e., a moving group.

\subsection{Star formation history of the region }
\label{sec:SFH}

 At least two subgroups of clusters can be identified in HC8. The age subgroups include NGC~7429, Stock 12, and Theia~391 with log \textit{t$\approx$8}, and the remaining clusters with log \textit{t$\approx$7 }(Table~\ref{tab:table1}). These subgroups suggest at least two distinct star formation events. It is noteworthy that all the likely members are relatively young ($\mathrm{<}$ 150 Myr), despite age not being a requisite in our selection criteria.~

 The relative proximity of Alessi 20, HSC~899 and Duvia 1 in terms of position, PM, \textit{plx}, and age (CMD, see below) suggests that at least these clusters were formed in a single event. They appear to radiate from a common origin, with their tangential velocities seemingly proportional to their distance from it (Figure~\ref{fig:image3}). HR24 provides median estimates of $\log t$  of 6.83 and 6.84 (ca. 7 Myr) for two of these clusters and we have obtained a consistent value for Duvia 1: 6.9. 

  {We performed backward orbital integration for all the eleven clusters of the OSC HC8 using the \textit{galpy} package \citep{2015ApJS..216...29B} in the Milky Way potential. We employed their present-day characteristics as in Table~\ref{tab:table1} for the starting point of Monte-Carlo sampling with Gaussian errors of 100 synthetic ``clusters''. The sampling was performed over \textit{plx}, PM and RV. The observational errors used as standard deviations were taken from HR24 and (if obtained in this work) from Table~\ref{tab:table1}. 
 Further, similarly to \cite{2024Natur.631...49S}, we have
generate 1000 bootstrapped samples by randomly drawing clusters and calculating evolution of the characteristic ``size'' of the supercluster as the median distance of clusters to its center, along with its uncertainty.}

\begin{figure}
 \begin{center}
\includegraphics[width=0.95\linewidth]{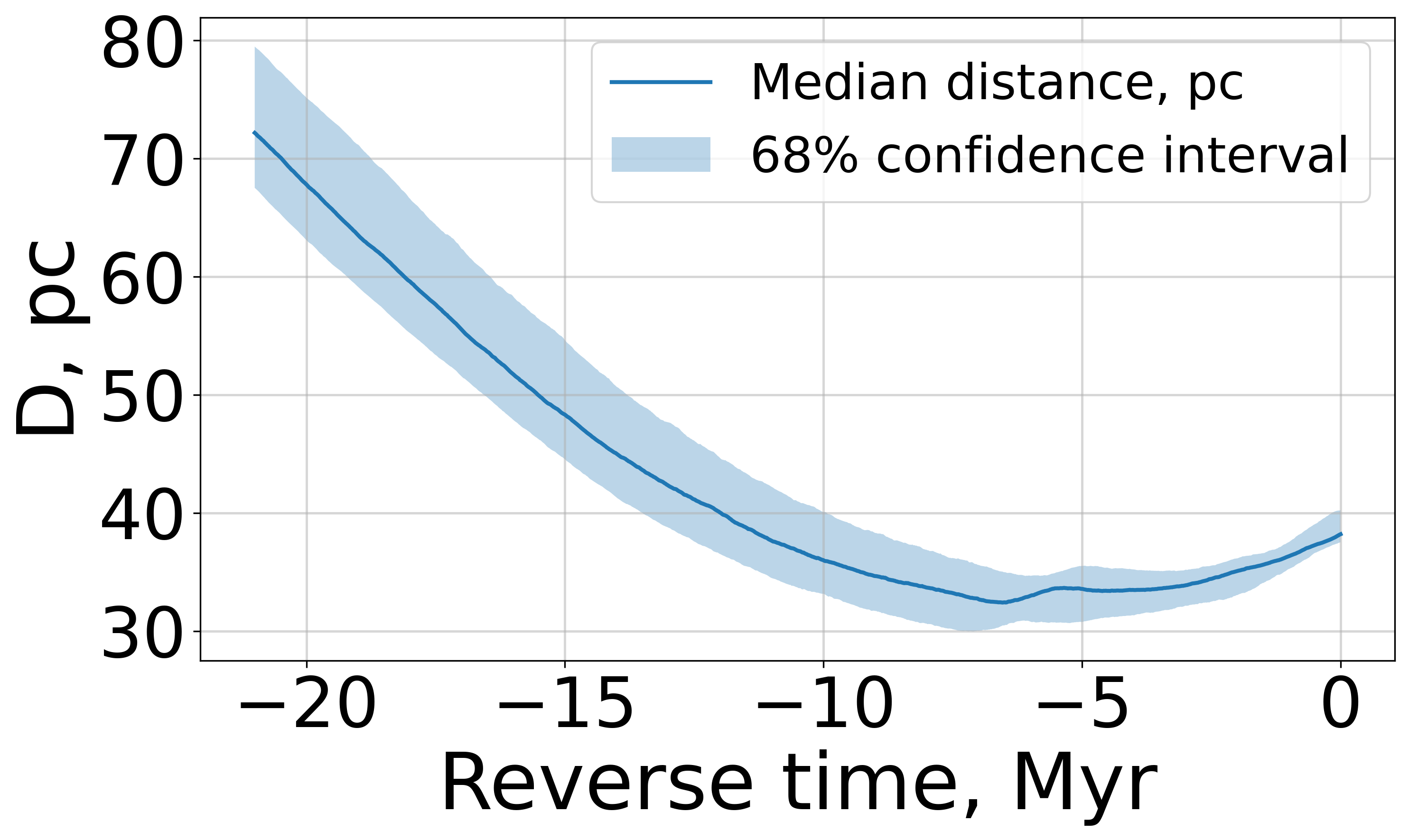}
\end{center}
\caption{ {Evolution of the characteristic size of the supercluster HC8 at backward integration of clusters' orbits in MW potential. Time=0 corresponds to present time. The space filled with light blue represents $68\%$ confidence interval.}}
\label{fig:image9}
\end{figure}

Figure~\ref{fig:image9} shows how the median distance of the eleven clusters (Table~\ref{tab:table1}) from the geometric center of  HC8 evolved backward from now (Time = 0) to 20 Myr ago (Time = -20). Notably, there is a distinct minimum at approximately -7 Myr, fitting the CMD ages for the above cited subgroup of young clusters, and demonstrating that they were significantly closer to each other at the time of their formation. Moreover, the figure shows that all the eleven clusters were closer to each other at that time.

\begin{figure}
 \begin{center}
\includegraphics[width=0.89\linewidth]{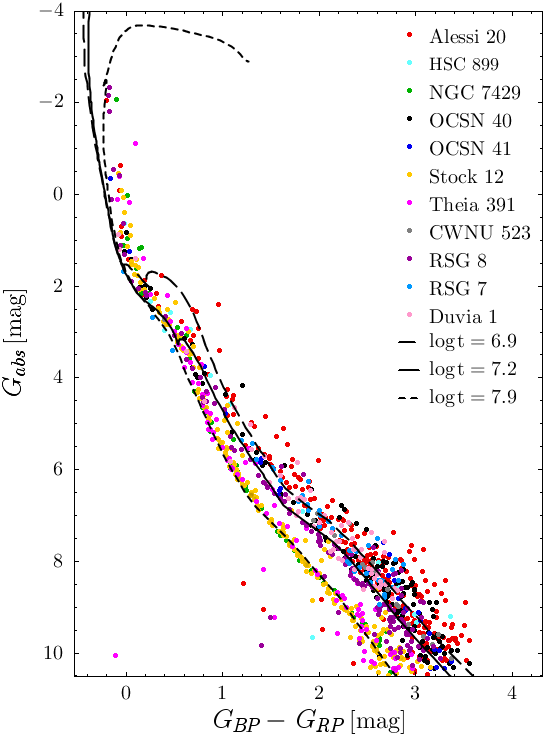}
\end{center}
\caption{The CMD of the eleven clusters in the supercluster HC8 and the fitting Padova isochrones (http://stev.oapd.inaf.it/cgi-bin/cmd) by \cite{2012MNRAS.427..127B} for \textit{Z}=0.0152. To convert to absolute magnitudes and dereddened colors we use estimates of interstellar extinction for clusters by HR24. For Duvia~1, interstellar extinction is from this work.}
\label{fig:image10}
\end{figure}

 Figure~\ref{fig:image10} represents the CMDs of clusters forming the supercluster, along with the isochrones for solar metallicity for three age groups obtained for these clusters. The observed "three CMDs" can be explained by three generations of clusters roughly 6-10 Myr, 13-17 Myr, and $\mathrm{\sim}$100 Myr old.  \cite{2022ApJ...931..156P} found a concordant span of ages (9 to $\mathrm{\sim}$100 Myr) in their Alessi 20 region. A somewhat similar situation is observed in the Vela region, where there are clusters of at least two different generations ($\mathrm{\sim}$5-10 Myr and $\mathrm{\sim}$30-50 Myr) \citep{2019A&A...626A..17C}.  {Another example of diverse ages spanning ca. 100 Myr within a group of clusters sharing a common kinematic signature has recently been reported in Cas OB5 \citep{2025A&A...697A..47Q}.}

 To constrain the star formation rate in the vicinity of the supercluster, we searched for massive O- and early B-type stars (earlier than $\mathrm{\sim}$B1V) within a distance of 20 pc from the centers of the constituent clusters. Such stars are indicators of recent star formation episodes, as their lifetimes are at most  {a few} Myr \citep{Maeder2000}. The catalog by \cite{2021A&A...650A.112Z} indicates that there are only two instances of such stars within a 20-pc radius centered on two of the clusters, RSG~7 and Duvia~1. Both clusters have the same early-type star (Gaia DR3 2013382713057701248) in their proximity. Extending the search radius to 50 pc does not significantly change the overall picture. Only one more early-type star is recovered, which could be associated with several of the clusters studied in this work (Table~\ref{tab:table2}).

 When we searched the StarHorse catalog for the presence of stars with surface temperatures above 30,000 K within 20 pc of the studied clusters, we identified none. Therefore, the region of the supercluster appears to have been inactive for several Myr.

\begin{table*}[tbp]
\label{tab:table2}
\begin{center}
\caption{O and early-B stars within 50 pc of any of the clusters forming the supercluster HC8.}
\setlength{\tabcolsep}{3pt}
\renewcommand{\arraystretch}{1.2}
\begin{tabular}{cccccc}
\hline 
Gaia DR3 source\_id & l & b & G & (BP-RP) & Clusters within 50 pc \\
                & (deg) & (deg) & (mag) & (mag) & \\
\hline
1993966437221117568 & 115.555 & -6.364 & 4.983 & -0.112 & OCSN~40, OCSN~41, CWNU~523 \\
2013382713057701248 & 109.948 & -0.783 & 4.792 & -0.031 & NGC~7429, HSC~899, RSG~8, RSG~7, Duvia~1 \\
\hline
\end{tabular}
\end{center}
\end{table*}

 There are several HII regions in the studied field (e.g., IC 1470, LBN 108.96+01.70, LBN 111.14-00.72, SHSH 2-161),  {according to the Simbad database,} some of which are associated with clusters (e.g., SH2-142 with NGC~7380, GAL 112.24+00.23 with [BDS2003] 44). However, all these objects are much farther away than HC8. Therefore, there are no prominent remnants of the GMC that formed this OSC.

\section{Concluding remarks}
\label{sec:concl}

 The scenario arising from our results suggests that a preceding GMC began star formation approximately 100~Myr ago and experienced successive starbursts until a few Myr ago.  The original clouds and OB stars have almost disappeared due to residual gas expulsion and supernova explosions, but at least 12 groups containing more than 1200 stars remain in the region, attesting to this history. Two of these star aggregates (Duvia~1 and a minor group) and the entire supercluster HC8  have been identified and characterized for the first time (Table~\ref{tab:table1}).

 {HC8 would be the outcome of a hierarchically structured SFR where age spreads are expected.} The CMD of all the stars shows three different main sequences, which can be explained by three generations of clusters, roughly 8, 15, and 100 Myr old. The photometric age of at least Alessi 20, HSC~899 and Duvia 1 matches their dynamical age, indicating a common origin of this subgroup from a relatively small volume. The whole supercluster was also more compact in the past.
We have demonstrated that the number of stars is a good proxy for mass estimation and that the mean mass of the stars in HC8 is close to solar mass.

 The system, along with five other open superclusters (OSCs), including a new one, HC9, are less than 500 parsecs from the Sun. Some of them have associated  {known} SFR and/or OB associations, but this is not the case for HC8, confirming that OSCs can last a few hundred Myr, despite generally dispersing within 150 Myr \citep{2024A&A...687A..52C}.  {Primordial groups are defined not just by age but by shared origin. Even if some clusters are not so young, they may still trace the same GMC; exhibit kinematic streaming and reside in coherent spatial structures.} 
 Although some of the clusters studied contain diverse subclusters and several stars have been attributed to two different clusters, no binary clusters have been found in this study. As with the other OSCs we studied \citep{2023MNRAS.521.1399C, 2024A&A...687A..52C}, the entire system is dispersing, and only three bound clusters (Stock 12, Alessi 20, and RSG~8) are expected to persist after a few hundred Myr. This is unsurprising: if OSCs were stable, many more of them (and older ones) should have been detected.

\begin{acknowledgements}

This work has made use of data from the European Space Agency (ESA) mission {\it Gaia} (\url{https://www.cosmos.esa.int/gaia}), processed by the {\it Gaia} Data Processing and Analysis Consortium (DPAC, \url{https://www.cosmos.esa.int/web/gaia/dpac/consortium}). Funding for the DPAC has been provided by national institutions, in particular the institutions participating in the {\it Gaia} Multilateral Agreement. The use of TOPCAT, an interactive graphical viewer and editor for tabular data \citep{2005ASPC..347...29T}, is acknowledged.

\end{acknowledgements}

\bibliographystyle{aa}  
\bibliography{main}

\newpage
\onecolumn
\appendix
\section{Census of the cluster Duvia~1}
\renewcommand{\arraystretch}{1.05}
\begin{longtable}{llllllllll} \\
\caption{The list of probable members of Duvia~1.}   
\label{tab:A} \\
\hline
\multicolumn{1}{l}{source\_id} &
\multicolumn{1}{l}{l  } &
\multicolumn{1}{l}{b } &
\multicolumn{1}{l}{plx }& 
\multicolumn{1}{l}{$\mu_\alpha$ } &
\multicolumn{1}{l}{$\mu_\delta$} &
\multicolumn{1}{l}{$G$} &
\multicolumn{1}{l}{(BP-RP)} &
\multicolumn{1}{l}{RV} &
\multicolumn{1}{l}{ruwe} \\
\multicolumn{1}{l}{} &
\multicolumn{1}{l}{(deg) } &
\multicolumn{1}{l}{(deg)} &
\multicolumn{1}{l}{(mas)} &
\multicolumn{1}{l}{(mas/yr)} &
\multicolumn{1}{l}{(mas/yr)} &
\multicolumn{1}{l}{(mag)} &
\multicolumn{1}{l}{(mag)} &
\multicolumn{1}{l}{(km/s)} &
\multicolumn{1}{l}{} \\
\hline
  2013250462421905408 & 108.94244771 & -1.63109416 & 2.288 & 6.584 & -1.803 & 16.73 & 2.66 & -- & 2.18\\
  2009291464648555648 & 110.64341283 & -3.47314306 & 2.361 & 6.462 & -1.839 & 16.85 & 2.92 & -- &0.98\\
  2009298023063556480 & 110.26307125 & -3.42306482 & 2.270 & 6.274 & -1.783 & 16.94 & 2.84 & -- &0.92\\
  2009298027358203648 & 110.26362630 & -3.42354865 & 2.272 & 6.283 & -1.977 & 16.56 & 2.61 & -- & 1.04\\
  2009579055657132800 & 110.35413947 & -2.70210694 & 2.266 & 5.987 & -1.640 & 15.85 & 2.34 & -- & 1.00\\
  2009586099404287232 & 110.84049509 & -2.62331351 & 2.319 & 6.501 & -1.626 & 17.43 & 3.09 & -- &0.98\\
  2009305586497384960 & 110.50194804 & -3.39612206 & 2.290 & 6.188 & -1.940 & 16.42 & 2.43 & -- & 1.15\\
  2009588500282206720 & 110.73334599 & -2.62286134 & 2.235 & 6.213 & -1.490 & 17.81 & 3.27 & -- & 1.08\\
  2009591803116118656 & 110.75766435 & -3.19042065 & 2.274 & 6.076 & -1.848 & 13.82 & 1.61 & -7.4 & 1.73\\
  2009602317196395264 & 111.30274607 & -3.07190442 & 2.304 & 6.117 & -2.006 & 16.32 & 2.79 & -- & 1.20\\
  2009607849114206336 & 111.31065027 & -2.92346148 & 2.361 & 6.444 & -1.707 & 16.26 & 2.49 & -- &0.97\\
  2009642930407684736 & 111.88767333 & -2.74320472 & 2.322 & 6.612 & -1.734 & 15.28 & 2.09 & -55.7 &0.98\\
  2009127186434989184 & 109.01224387 & -3.23821503 & 2.276 & 6.166 & -1.665 & 14.89 & 1.92 & 13.3 & 2.70\\
  2009131382628063488 & 108.97880305 & -3.12489538 & 2.286 & 5.786 & -2.112 & 17.45 & 3.03 & -- & 1.07\\
  2009344481721590912 & 111.22799615 & -3.80711624 & 2.303 & 6.632 & -2.073 & 15.00 & 2.26 & -- & 1.07\\
  2009149146613629312 & 109.56042787 & -3.16450285 & 2.344 & 6.501 & -1.782 & 15.74 & 2.29 & -- & 1.04\\
  2009696669038165120 & 111.46006114 & -2.63086850 & 2.249 & 6.651 & -1.956 & 15.68 & 2.39 & -- & 2.18\\
  2009408867573057024 & 111.26095615 & -3.25695362 & 2.337 & 6.457 & -1.829 & 17.63 & 3.13 & -- & 1.09\\
  2009177699552606848 & 109.76127089 & -4.23661303 & 2.334 & 6.259 & -2.080 & 16.14 & 2.54 & -- & 1.01\\
  2009438623105192064 & 111.41272144 & -3.05885597 & 2.322 & 6.579 & -2.115 & 16.89 & 2.85 & -- & 1.05\\
  2009467111627769856 & 110.12065868 & -3.30202657 & 2.243 & 5.998 & -1.826 & 14.78 & 1.78 & -1.3 & 1.06\\
  2009490441890468736 & 110.29014815 & -3.28558274 & 2.328 & 6.265 & -1.474 & 17.32 & 3.03 & -- & 1.02\\
  2009498374698931712 & 110.48877271 & -3.23951602 & 2.249 & 6.110 & -1.559 & 17.72 & -- & -- &0.95\\
  2009504048346789120 & 110.20928286 & -3.17105245 & 2.264 & 5.790 & -1.881 & 10.08 &0.40 & -- &0.92\\
  2009252260186167424 & 109.80353555 & -3.78203659 & 2.322 & 6.001 & -2.029 & 17.72 & 3.10 & -- & 1.12\\
  2009536655739544576 & 110.19787494 & -2.94112614 & 2.352 & 6.282 & -1.846 & 14.13 & 1.77 & -18.3 & 1.18\\
  2009268370601119360 & 110.07206036 & -3.60669752 & 2.231 & 5.909 & -1.820 & 16.98 & 2.63 & -- & 1.06\\
  2009560290949511040 & 110.27450180 & -2.91686082 & 2.301 & 5.920 & -1.653 & 16.37 & 2.71 & -- & 1.90\\
  2010077237508787328 & 108.66032913 & -2.23427418 & 2.268 & 5.807 & -1.850 & 17.00 & 3.16 & -- & 1.11\\
  2010099159022859392 & 108.14338032 & -2.11420138 & 2.242 & 6.090 & -1.485 & 16.22 & 2.25 & -- & 1.05\\
  2009909218383022464 & 109.21336100 & -2.92639897 & 2.209 & 6.303 & -1.681 & 15.89 & 2.26 & -- & 1.14\\
  2010305042566720640 & 110.08934015 & -2.67197086 & 2.331 & 6.277 & -1.872 & 16.63 & 2.83 & -- & 1.76\\
  2010315659725644160 & 109.93326132 & -2.52748063 & 2.235 & 6.073 & -1.634 & 17.29 & -- & -- &0.98\\
  2010316415639655040 & 109.78634871 & -2.56455190 & 2.310 & 5.973 & -1.574 & 17.78 & 3.29 & -- & 1.09\\
  2010321913198046848 & 110.07779780 & -2.51371059 & 2.295 & 5.784 & -1.519 & 17.67 & 3.17 & -- &0.92\\
  2010322222435670272 & 110.09752747 & -2.49325478 & 2.209 & 6.253 & -2.023 & 17.44 & 2.85 & -- & 1.04\\
  2010347820441541120 & 110.43998159 & -2.44839595 & 2.331 & 5.909 & -1.511 & 17.50 & 3.06 & -- & 1.06\\
  2010358983056051328 & 110.86372015 & -2.28952124 & 2.209 & 6.047 & -1.728 & 17.16 & 2.92 & -- & 1.03\\
  2010367474212803968 & 109.98659046 & -2.39816337 & 2.299 & 6.476 & -1.878 & 17.15 & 2.72 & -- & 1.03\\
  2010376025489405824 & 110.24287906 & -2.29706402 & 2.289 & 5.874 & -1.938 & 17.99 & 3.24 & -- &0.99\\
  2010383623289536768 & 109.83122826 & -2.21173195 & 2.207 & 6.072 & -1.876 & 17.19 & 2.68 & -- & 1.03\\
  2010407503308555776 & 110.66296615 & -2.09096330 & 2.271 & 5.852 & -1.647 & 10.59 &0.59 & -23.5 &0.92\\
  2010188356897983360 & 109.30804246 & -1.92776028 & 2.256 & 5.777 & -1.700 & 17.75 & 3.05 & -- &0.96\\
  2010443168714897408 & 111.11616358 & -2.27154265 & 2.329 & 6.209 & -1.712 & 16.51 & 3.02 & -- & 1.05\\
  2010766047179013760 & 350.70195751 & 59.65747434 & 2.276 & 6.113 & -1.349 & 10.39 & 0.41 & -20.34 & 0.84 \\
  2010938563121841920 & 112.28663905 & -3.01694874 & 2.313 & 6.165 & -1.900 & 12.57 & 1.11 & -9.4 &0.90\\
  2013725451443983872 & 110.72032924 &-0.94599653 & 2.229 & 5.868 & -1.560 & 17.02 & 2.60 & -- & 1.70\\
  2013540866630422656 & 109.33848037 &-0.77215682 & 2.260 & 5.788 & -1.809 & 17.72 & 2.81 & -- & 1.04\\
  2013367594772877568 & 109.98674163 &-0.93096763 & 2.206 & 5.973 & -1.659 & 17.86 & 2.95 & -- &0.97\\
  1997260397954832000 & 111.15101613 & -4.29561098 & 2.278 & 6.587 & -2.085 & 17.61 & 3.12 & -- & 1.07\\
  1997292562972794496 & 110.97322356 & -4.10892499 & 2.217 & 6.376 & -1.974 & 16.15 & 2.27 & -- & 1.11\\
  1997303936046069504 & 110.86191161 & -4.01929650 & 2.251 & 6.297 & -2.022 & 15.86 & 2.27 & -- &0.99\\
  1997329053015500672 & 111.16168248 & -3.99478096 & 2.331 & 6.310 & -1.835 & 15.76 & 2.17 & -- & 1.03\\
\hline
\end{longtable}



\end{document}